\documentclass[a4paper,11pt]{article}
\usepackage{pos}
\usepackage{subfigure}  

\usepackage{graphicx}
\usepackage{dcolumn}
\usepackage{bm}

\title{Source models of ultrahigh-energy cosmic rays}

\author*[a,b]{Bing Theodore Zhang}

\affiliation[a]{Key Laboratory of Particle Astrophysics, Institute of High Energy Physics, Chinese Academy of Sciences, Beijing 100049, China}
\affiliation[b]{TIANFU Cosmic Ray Research Center, Chengdu, China}

\emailAdd{zhangbing@ihep.ac.cn}

\abstract{We investigate potential sources of ultrahigh-energy cosmic rays (UHECRs) and their acceleration mechanisms, focusing on astrophysical phenomena associated with massive stellar deaths and supermassive black holes. These phenomena include gamma-ray bursts (GRBs), engine-driven supernovae/hypernovae, magnetars, newly born pulsars, binary neutron star mergers (BNS), tidal disruption events (TDEs), and active galactic nuclei (AGN). While high-luminosity GRBs (HL GRBs) are constrained as UHECR sources by high-energy neutrino observations, low-luminosity GRBs (LL GRBs) and engine-driven supernovae remain promising candidates, with intermediate-mass nuclei as the dominant components. Compact binary mergers and $r$-process nucleosynthesis in neutron-rich environments may also contribute to ultraheavy UHECRs.
The composition of UHECRs from TDEs depends on the properties of the disrupted stars. AGN, particularly radio galaxies, remain promising sources, with acceleration occurring in their large-scale jets and lobes. Shear acceleration mechanisms have been proposed as a viable alternative for accelerating UHECRs, involving the re-acceleration of low-energy cosmic rays and being compatible with the observed spectrum and composition.
Future multi-messenger observations, especially from upcoming observatories, are expected to provide critical data to refine our understanding of UHECR origins, test existing models, and explore new acceleration mechanisms.}

\FullConference{7th International Symposium on Ultra High Energy Cosmic Rays (UHECR2024)\\
 17-21 November 2024\\
Malargüe, Mendoza, Argentina\\}

\begin{document}
\maketitle

\section{Introduction}
Ultrahigh-energy cosmic rays (UHECRs) are relativistic charged nuclei originating from space, with energies exceeding the EeV threshold. The origin of UHECRs remains one of the most significant unresolved problems in modern astroparticle physics \citep{Kotera:2011cp}. Since their first detection in the 1960s \citep{Greisen:1960wc}, substantial progress has been made in measuring their energy spectrum. 
Significant advances have also been achieved in determining the composition of UHECRs, which is crucial for understanding their origins \citep{PierreAuger:2022atd}. The large-scale anisotropy of UHECRs, particularly the dipole observed by the Auger Collaboration, provides strong evidence that UHECRs originate from extragalactic sources \citep{PierreAuger:2017pzq}. 
This suggests that UHECR sources are correlated with the local distribution of matter in the Universe.

In the following sections, we summarize the latest observations of UHECRs, discuss methods to infer the properties of their sources, examine potential astrophysical candidates, and provide a concluding summary and perspective.

\section{Latest observations}
Around an energy of $\sim 4\rm~EeV$, there is an ``ankle'', marking the transition from Galactic to extragalactic origins\footnote{The transition could also occur at lower energy.}.
At an energy of around $\sim 10\rm~EeV$, there is a feature known as the ``instep".
There is also a steep flux suppression beyond $\sim 40\rm~EeV$~\cite{PierreAuger:2020qqz, PierreAuger:2020kuy}.  
The suppression could be due to interactions with CMB/EBL photons, known as the GZK cutoff~\cite{Greisen:1966jv, Zatsepin:1966jv}, or it may reflect the maximum acceleration energy at the sources~\cite{Kotera:2011cp}.
However, a discrepancy has been observed between the measured spectra of Auger and the Telescope Array (TA), which may be related to differences in energy scale measurements~\cite{PierreAuger:2023wti}.
Recently, TA detected the Amaterasu particle with an energy of $E = 240\rm~EeV$, the second-highest-energy cosmic ray detected on Earth~\cite{TelescopeArray:2023bdy}.
The highest-energy UHECR, at 320 EeV, was detected by the Fly's Eye air shower detector in 1991~\cite{HIRES:1994ijd}.
However, no clear sources have been identified for these highest-energy CRs.

The depth of the cosmic-ray shower maximum, $X_{\rm max}$, is a measurable quantity used to infer the particle composition~\cite{Greisen:1960wc}.
Auger data suggest a mixed composition of UHECRs, with intermediate-mass nuclei(e.g., carbon and oxygen) and/or heavy nuclei (e.g., iron) contributing significantly beyond $10\rm~EeV$~\citep{PierreAuger:2023xfc}.
In particular, the fraction of protons gradually decreases above the ankle, 
while intermediate-mass nuclei may become dominant at higher energies. 
The contribution of heavy nuclei appears negligible in the energy range of $10^{18.4}-10^{19.4}\rm~eV$~\cite{PierreAuger:2023xfc}, but these results are strongly influenced by hadronic interaction models. 
The distribution of $X_{\rm max}$, 
measured by TA is consistent with Auger data, although the interpretation is still debated~\cite{PierreAuger:2023yym}.
Recently, the TA collaboration has studied the mass composition based on the arrival directions of UHECRs. 
They found that the composition is consistent with heavy nuclei.
Furthermore, the observed isotropy of UHECRs above $100\rm~EeV$ suggests that their composition is even heavier, possibly exceeding iron nuclei~\cite{TelescopeArray:2024oux, TelescopeArray:2024buq}.

The measurement of the dipole is one of the key achievements in UHECR detection.
The significance of the dipole is currently around $6.8\sigma$ above $8\rm~EeV$~\cite{PierreAuger:2024fgl}. 
On intermediate scales, a hot spot has been observed in the Centaurus constellation with a significance of $4\sigma$ for UHECRs with energies above $8\rm~EeV$. 
Potential sources in this region include the radio galaxy Cen A and the starburst galaxy NGC 4945.
The correlation with starburst galaxies (SBGs) has a significance of $4.7\sigma$~\cite{Salamida:2023fmk}.
The TA collaboration has also observed a hot spot with a significance of $3.2\sigma$~\cite{TelescopeArray:2023bdy}, as well as an excess in the Perseus-Pisces supercluster with the same significance of $3.2\sigma$~\cite{TelescopeArray:2023bdy}.
More statistics are needed to fully understand the source properties of UHECRs.

\section{Implications of the sources of UHECRs}
Where do UHECRs originate? Traditionally, potential sources have been inferred based on their acceleration capabilities and energy budgets. In recent years, however, we can also explore the source properties more effectively by studying the observed compositions and anisotropy. Additionally, the rigidity distribution of UHECRs provides further constraints on their sources.

\subsection{Hillas condition and energetics of UHECRs}
The Hillas condition is often considered a confinement condition, representing the maximum acceleration capability of a source~\cite{Hillas:1985is}. 
Assuming the maximum magnetic luminosity is smaller than the bolometric luminosity, $L_B \lesssim L_{\rm bol}$, the required minimum bolometric luminosity is 
\begin{align}
    L_{\rm bol, min} &> \frac{1}{2} \Gamma^2 \beta c \left(\frac{E}{Ze\beta}\right)^2 \sim 10^{45}\Gamma^2 \beta^{-1} \left(\frac{E}{Z 10^{20}\rm~eV}\right)^2,
\end{align}
where $Z$ is the nuclear charge, $\Gamma$ is the Lorentz factor, and $\beta$ is the dimensionless velocity of the source.
Various sources meet this condition, making them potential candidates for UHECRs. 
These sources include energetic transients in star-forming galaxies, such as trans-relativistic supernovae/hypernovae~\cite{Murase:2006mm, Wang:2007ya, Murase:2008mr, Chakraborti:2010ha, Zhang:2017moz}, long and short gamma-ray bursts (GRBs)~\cite{Waxman:1995dg, Vietri:1995hs, Zhang:2024sjp}, and binary neutron star mergers (BNS)~\cite{Takami:2013rza, Kimura:2018ggg, Rodrigues:2018bjg}. 
Magnetars and young pulsars have also been proposed as possible sources~\cite{Fang:2012rx}.
High-energy sources related to supermassive black holes are powerful accelerators, including active galactic nuclei (AGN)~\cite{Takahara1990, Rachen:1992pg, Murase:2011cy} and tidal disruption events (TDEs)~\cite{Farrar:2008ex, Farrar:2014yla, Zhang:2017hom}.
Merger shocks within galaxy clusters~\cite{Norman1995, Kang:1995xw} and galactic superwinds of starburst galaxies (SBGs)~\cite{Jokipii1985, Anchordoqui:1999cu} could also accelerate UHECRs. 
An open question is whether the sources of UHECRs share similar properties. 
Assuming the maximum energy depends on luminosity, $R_{\rm max} \propto (L/L_0)^{1/2}$, it is easy to see that the rigidity follows a power law distribution: $p(R_{\rm max}) \propto R_{\rm max}^{-\beta_{\rm pop}}$ for a given luminosity function.
Recent analysis suggests narrow rigidity and nearly identical sources, with a factor of 2 dispersion in maximum rigidity~\cite{Ehlert:2022jmy}.
However, acceleration mechanisms may not directly reflect the luminosity function.
This conclusion depends on the source models, including where and how the acceleration occurs.

\begin{figure}
    \centering
    \includegraphics[width=1.0\linewidth]{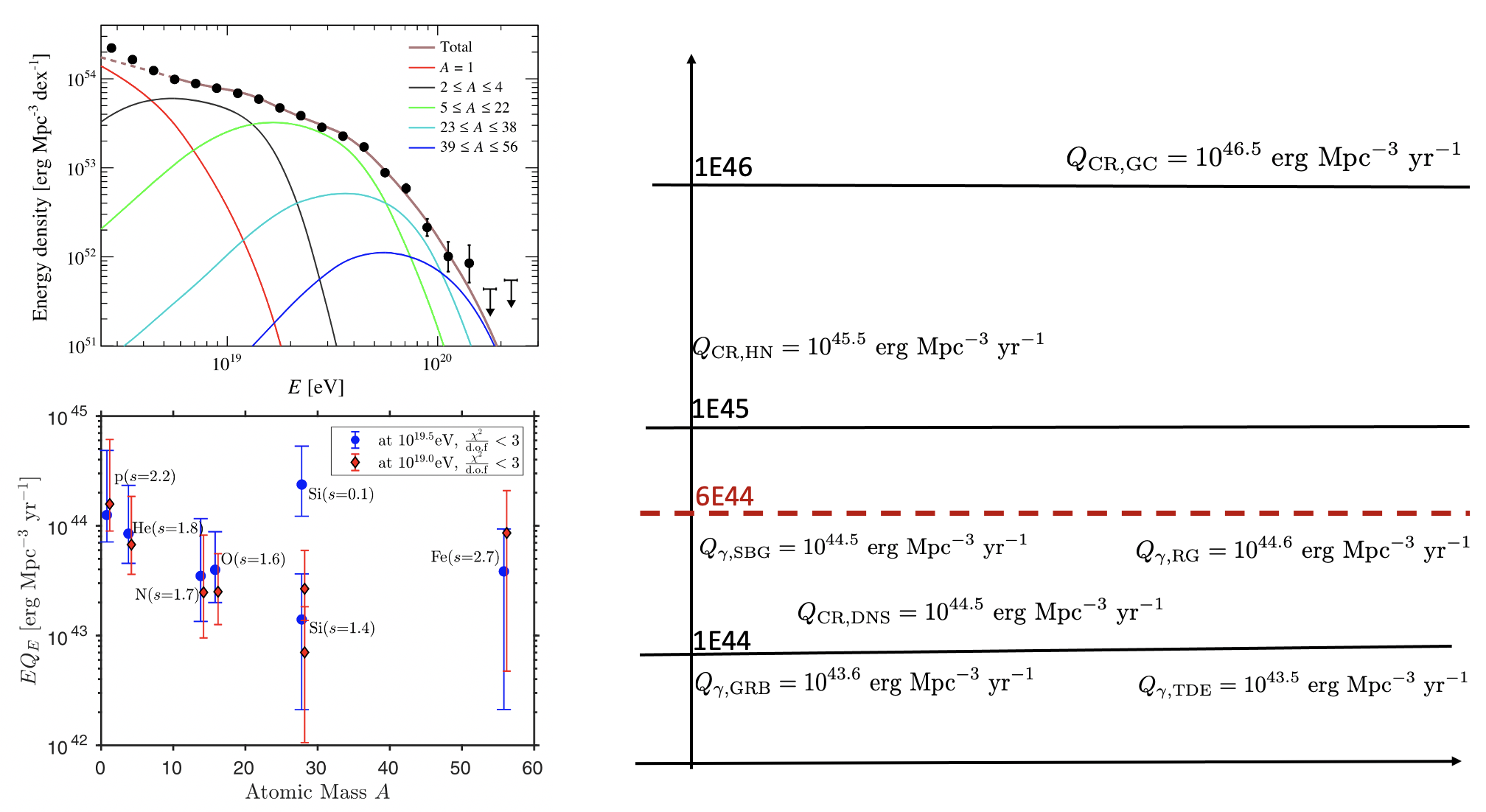}
    \caption{Energy injection rate of UHECRs and energetics of potential sources~\cite{PierreAuger:2020kuy, Jiang:2020arb}.}
    \label{fig:UHECR-energetic}
\end{figure}

The total energy generation density of UHECRs above $\sim 3\rm~EeV$, as inferred from observations, is $Q_{\rm inj} \sim 6 \times 10^{44}\rm~erg~Mpc^{-3}~yr^{-1}$ for mixed composition~\cite{PierreAuger:2020kuy}.
Ref.~\cite{Jiang:2020arb} studied the differential energy generation rate density for different species, finding $EQ_E^{19.5} \sim (0.4 - 1.5) \times 10^{44}\rm~erg~Mpc^{-3}~yr^{-1}$ for UHECRs with spectral index $s \gtrsim 1.5$ at $E = 10^{19.5}\rm~eV$, see the lower left panel in Fig.~\ref{fig:UHECR-energetic}.
We show the energetics of various UHECR sources in the right panel of Fig.~\ref{fig:UHECR-energetic}, where we can see that most of the candidate sources satisfy the required energetics.

\subsection{Composition is the key to identifying the sources of UHECRs}
The source properties can be inferred from the combined fit of the observed energy spectrum and composition~\cite{Aab:2016zth, PierreAuger:2022atd}. 
The best-fit results involve an injection ratio of $f_{\rm H} : f_{\rm He} : f_{\rm N} : f_{\rm Si} : f_{\rm Fe} = 0 : 24.5\% : 68.1\% : 4.9\% : 2.5\%$~\cite{PierreAuger:2022atd}.
We can see that UHECR source has a mixed composition. 
In particular, there is a large fraction of intermediate- and heavy-mass nuclei, while the proton component is very small.
At GeV energies, CRs are primarily protons, which is consistent with the solar composition~\cite{ParticleDataGroup:2024cfk}. 
The present-day solar composition for H, He, and metals is $X = 0.793$, $Y = 0.2469$, and $Z = 0.0141$, respectively, and the most abundant heavy nuclei are $\rm O$, $\rm C$, $\rm Ne$, and $\rm Fe$ \cite{Lodders:2010kx}.
However, the fitting results suggest that UHECR sources require a significant enhancement of heavier nuclei, and the solar composition does not explain the observed data~\cite{Zhang:2017hom}. 

For sources linked to the deaths of massive stars or compact stars, UHECRs could be accelerated in the ejecta or winds.
Intermediate- and heavy-mass nuclei could be drawn from stellar ejecta through the accretion disk via a ``one-time'' injection process, entrained through mixing with stellar ejecta or produced through explosive nucleosynthesis in the jets~\cite{ Metzger:2011xs}. 
Low-energy CRs could be further accelerated to the UHE energy range in the large-scale jet of radio galaxies~\cite{Kimura:2017ubz, Caprioli:2015zka}. 
Ref.~\cite{Kimura:2017ubz} considers the injection of low-energy CRs at TeV energies via the discrete shear acceleration mechanism considering further enhancement of heavy nuclei in radio galaxies.
Future precise measurements of the composition, especially the single-component spectrum, could confirm such a scenario.
Intermediate and heavy nuclei could be further enhanced during the acceleration process, with the enhancement scaling as $\propto (A / Z)^2$, especially for particles with a large charge-to-mass ratio~\cite{Caprioli:2017oun}.

\begin{figure}
    \centering
    \includegraphics[width=0.45\linewidth]{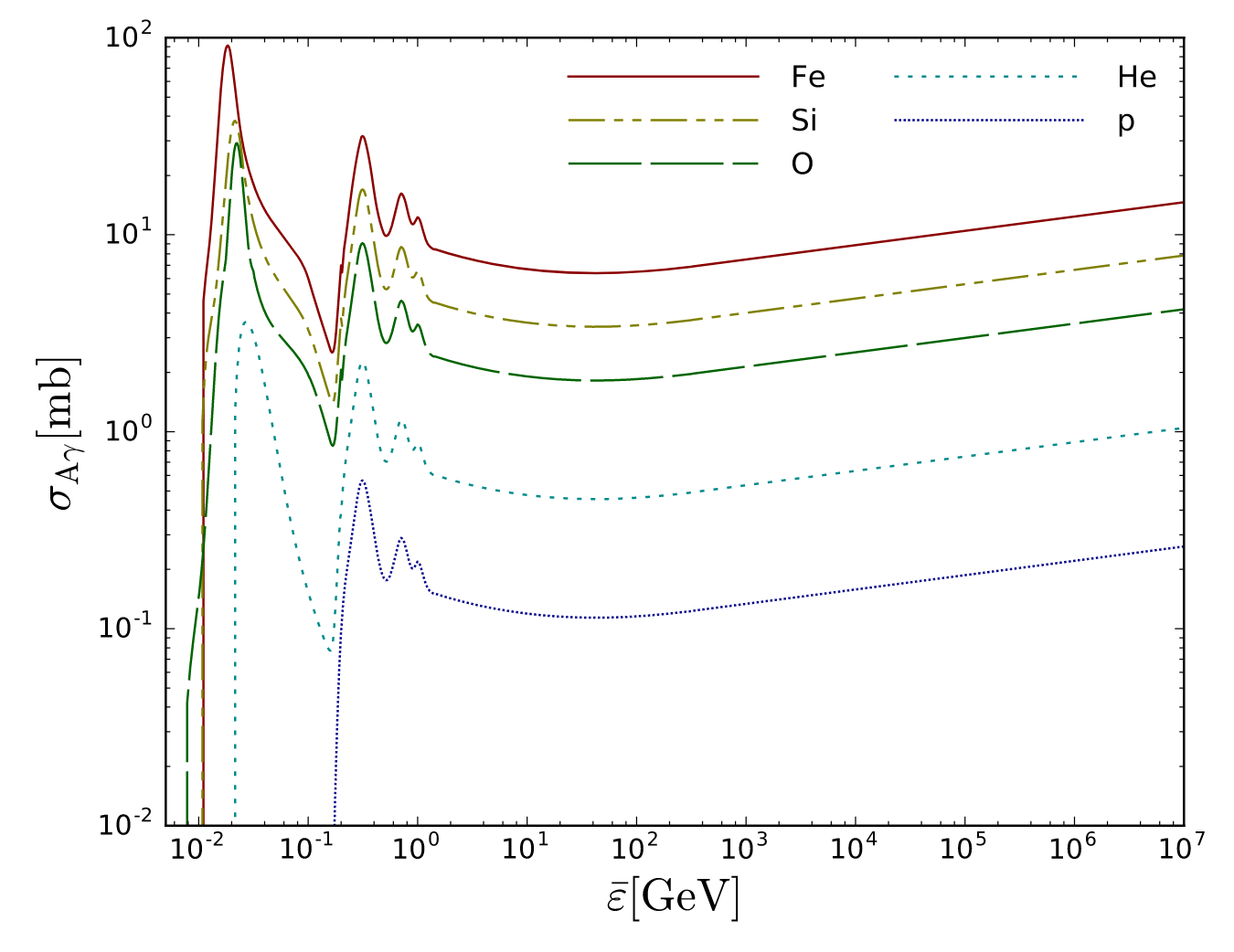}
    \includegraphics[width=0.47\linewidth]{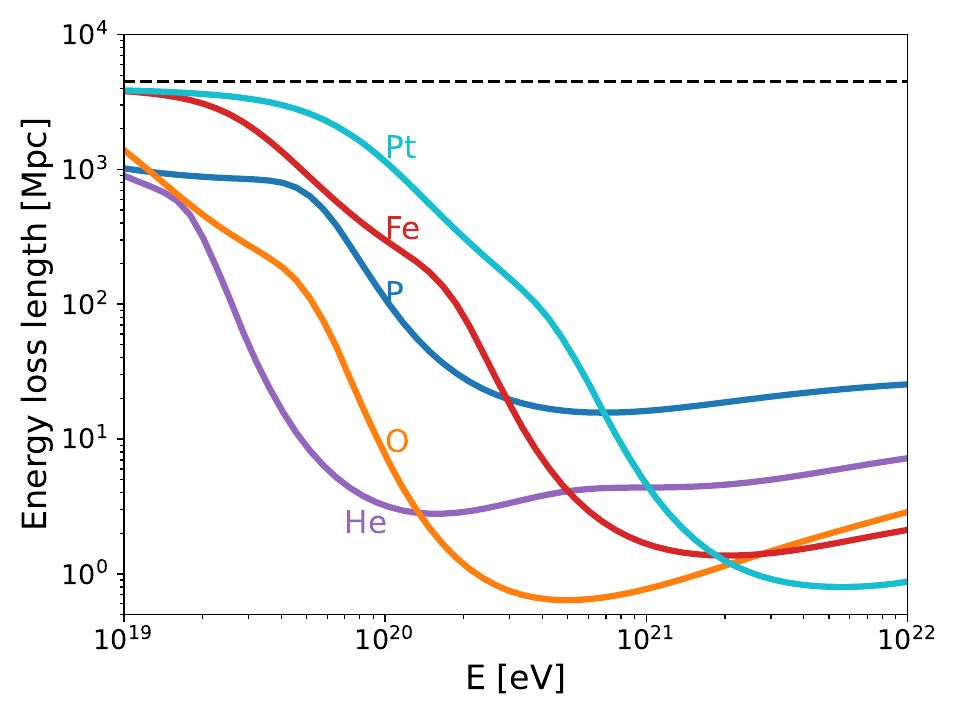}
    \caption{Right figure: Photonuclear and photomeson production cross sections for five chemical species. Right: Energy loss length of various nuclei, including ultraheavy UHECR nuclei~\cite{Zhang:2024sjp}.}
    \label{fig:cross-section}
\end{figure}

The survival of nuclei is one of the main arguments for identifing the sources of UHECRs. We show the photonuclear and photomeson production cross sections for five chemical species in Fig.~\ref{fig:cross-section}.
We can estimate the energy-loss timescale using the following formula,
\begin{equation}
t_{A\gamma}^{-1}(E_A) = \frac{c}{2 \gamma_A^2} \int_{\bar{\varepsilon}_{\rm th}}^{\infty} 
d\bar{\varepsilon} \sigma_{A\gamma}(\bar{\varepsilon})  \kappa_{A}(\bar{\varepsilon})\bar{\varepsilon}\int_{\bar{\varepsilon}/2\gamma_A}^{\infty} d\varepsilon \frac{1}{\varepsilon^2} \frac{dn}{d\varepsilon},
\end{equation}
where $\gamma_A$ is the Lorentz factor of UHECRs with mass number $A$, $\bar{\varepsilon}_{\rm th}$ is the threshold energy measured in the rest frame of the initial nucleus (NRF), and ${\rm d} n/{\rm d}\varepsilon$ is the differential number density of target photons~\cite{Zhang:2017hom}. Here, $\sigma_{A\gamma}(\bar{\varepsilon}) = A \sigma_{p \gamma}(\bar{\varepsilon})$ is the photohadronic cross section related to the photomeson or photodisintegration process. The inelasticity is $\kappa_{A\gamma}(\bar{\varepsilon}) \equiv \frac{\Delta E}{E} = \frac{\Delta N}{N}$, where $N$ is the total number of nucleons in the parent nucleus, and $\Delta N$ is the number of ejected nucleons in each channel. The survival of nuclei challenges luminous sources as the origin of the observed UHECR nuclei in the internal shock model, such as HL GRBs, TDEs, and luminous blazars. Note that the external reverse/forward shock model is not affected~\cite{Murase:2008mr, Zhang:2017hom, Zhang:2017moz}. 

\subsection{Sources inferred from anisotropy constraints}
The absence of small-scale anisotropy constrains the effective source number density, $N_s > 10^{-5} - 10^{-4}\rm~Mpc^{-3}$~\cite{PierreAuger:2013waq}, challenging ``rare'' steady sources~\citep{Takami:2014zva}.
Transients as UHECR sources could avoid these constrains~\cite{Murase:2008sa, Takami:2011nn}.
Recently, Ref.~\cite{Marafico:2024qgh} studied the anisotropy correlated with galaxies in the local sheet, considering transient sources in star-forming galaxies (or dwarfs).
The constraint on the burst rate density is $50 {\rm~Gpc^{-3}~yr^{-1}} < \rho_s < 30000 {\rm~Gpc^{-3}~yr^{-1}}$, which is consistent with long-duration GRBs or trans-relativistic supernovae/hypernovae.
The observed correlation with nearby starburst galaxies could be due to the echo of a past burst from Cen A~\citep{Bell:2021pkk, Taylor:2023qdy}.

\section{Candidate sources of UHECRs}
\subsection{Massive stellar deaths}
In recent years, there has been a surge in the detection of transient high-energy sources across the entire electromagnetic spectrum, from radio to gamma rays.
We consider transient sources linked to the deaths of massive stars,
including gamma-ray bursts (GRBs), engine-driven supernovae/hypernovae, pulsars, and magnetar-driven transients.

GRBs are the brightest and probably the most powerful high-energy astrophysical phenomena in the Universe~\cite{Meszaros:2006rc}.
High-luminosity GRBs (HL GRBs) have typical isotropic radiation luminosities of $\sim 10^{51} - 10^{53}\rm~erg~s^{-1}$.
GRBs are considered promising sources of UHECRs due to their acceleration capability and energetics~\cite{Waxman:1995dg, Vietri:1995hs}. 
However, the baryonic content of the GRB jet is strongly constrained by the absence of coincident detections of GRBs and high-energy neutrinos~\cite{IceCube:2012qza}. 
A detailed analysis of thousands of GRBs showed that the prompt neutrino emission from all GRBs in the Universe is limited to 1\% of the diffuse astrophysical neutrino flux~\cite{Abbasi:2022whi}.
In the future, multi-messenger observations of GRBs is critical to testing their ability to be the sources of UHECRs.
In contrast, low-luminosity GRBs could dominate diffuse neutrino flux detected by IceCube~\cite{Murase:2006mm}.

\begin{figure}
    \centering
    \includegraphics[width=1.0\linewidth]{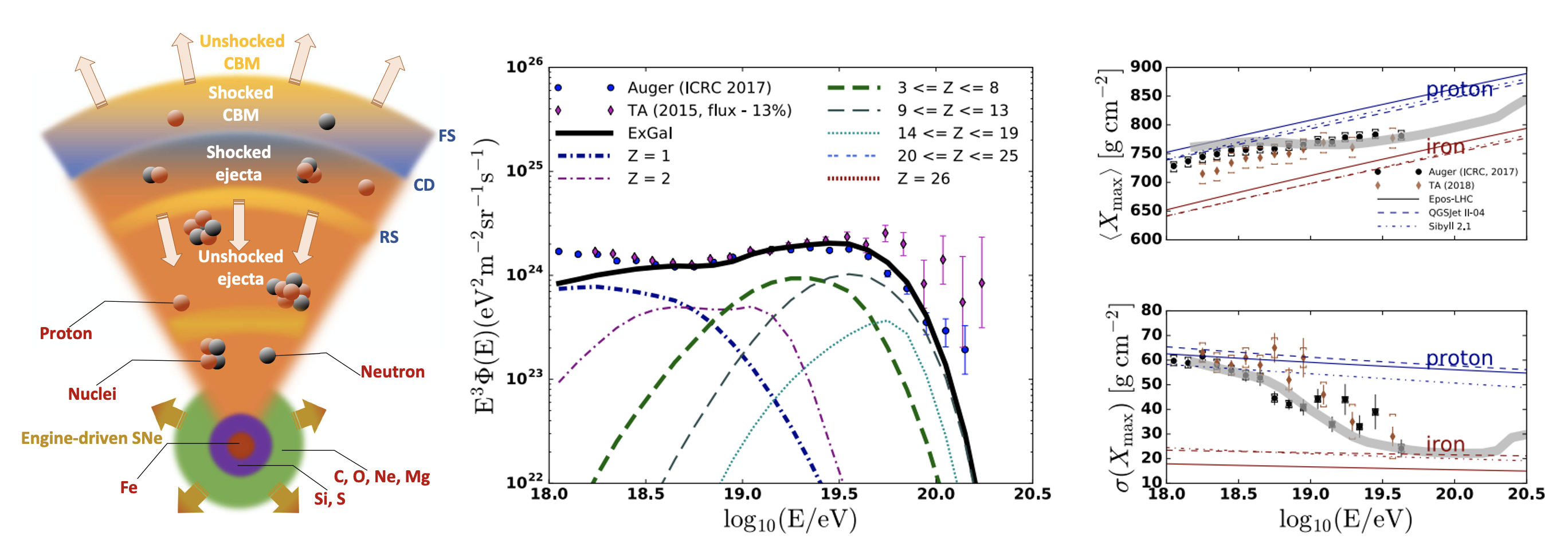}
    \caption{Left panel: A schematic diagram about the origin of UHECR nuclei
from GRBs. Right panel: The mean and standard deviation
of the measured Xmax distributions as a function of energy. Figure adapted from Ref.~\cite{Zhang:2017moz, Zhang:2018agl}.}
    \label{fig:MSD}
\end{figure}

Typically, UHECR nuclei cannot survive in the prompt emission region
In the afterglow phase, nuclei can survive for both HL and LL GRBs~\cite{Murase:2008mr, Zhang:2017moz}.
There are several reasons to consider LL GRBs as source of UHECRs, rather than HL GRBs.
First, unlike HL GRBs, LL GRBs allow nuclei to survive. 
Second, the narrow-rigidity problem could be alleviated if the external shock accelerates the CRs.
Third, the event rate of LL GRBs is higher than that of HL GRBs.
In Fig.~\ref{fig:MSD}, we show an example in which the relativistic outflow can extract nuclei from the stellar core and accelerate them to ultrahigh energies in the energy dissipation region via internal shocks or external reverse-forward shocks. 
We show the energy spectrum of the UHECR nuclei of the Jet-B model with mass fractions $X_{\rm O} = 0.511, X_{\rm Si} = 0.364$ and $X_{\rm S} = 0.108$, where the UHECR nuclei are in the complete survival regime. 
Newborn pulsars and magnetars are formed following the massive stellar collapse and low-mass neutron star merger remnant. 
Heavy nuclei may originate from the neutron star's surface and accelerate to the UHE energy range~\cite{Blasi:2000xm, Arons:2002yj, Murase:2009ah, Fang:2012rx}.
Current limits on the diffuse UHE neutrino flux from Auger and IceCube started to constrain the relevant parameter space~\cite{Fang:2013vla}.

\subsection{Compact binary mergers}
Compact binary mergers include binary neutron star mergers (NS-NS) and black hole-neutron star mergers (BH-NS).
Short GRBs are often associated with NS-NS and BH-NS mergers, and their jets are efficient acceleration sites for CRs, similar to those of long GRBs.
For merger shocks, the large charge number can enable ultra-heavy UHECR (UH-UHECR) nuclei to be accelerated to the UHE energy range.
UH nuclei are synthesized as a result of the r-process occurring in neutron-rich environments. 
The third-peak material exists in the equatorial plane (dynamical ejecta). 
Merger shocks from NS-NS or BH-NS mergers have been suggested as sources of UH-UHECR nuclei, which could be associated with the origin of the Amaterasu particle~\cite{Zhang:2024sjp, Farrar:2024zsm}.
Ref.~\cite{Zhang:2024sjp} performed a combined fit to Auger and TA data with UH-UHECR nuclei and found that the energy generation rate density is $Q_{\rm UH-UHECR}^{\rm Auger} \lesssim (0.1 - 15) \times 10^{42}\rm~erg~Mpc^{-3}~yr^{-1}$ for Auger data and $Q_{\rm UH-UHECR}^{\rm TA} \lesssim (1.4 - 5.6) \times 10^{43}\rm~erg~Mpc^{-3}~yr^{-1}$ for TA data.
This energy bound is consistent with the prediction for NS-NS or NS-BH events~\cite{Zhang:2024sjp}.

\subsection{Tidal disruption events}
Tidal disruption events are violent processes in which a star is disrupted by a central supermassive black hole (SMBH)~\cite{Rees:1988bf}.
TDEs have been suggested as sources of UHECRs~\cite{Farrar:2008ex, Farrar:2014yla, Zhang:2017hom, Guepin:2017abw, Piran:2023svv, Plotko:2024gop}.
A fraction of TDEs have relativistic jets, e.g., Sw J1644, Sw J2058, Sw J1112, and AT 2022cmc.
UHECRs could be accelerated in the internal and external reverse/forward shocks of TDE jets~\cite{Zhang:2017hom}.
Ref.~\cite{Piran:2023svv} showed that the maximum acceleration energy is achieved at hundreds of days after the optical discovery for off-axis jets considering the effect of sideways escape.
The energy output of jetted TDEs such as the off-axis AT2018hyz event is comparable to the required energy injection rate of UHECRs.

The composition of UHECRs is linked to the disrupted stars, where UHECRs are accelerated in either the internal shock or the external reverse shock.
Tidal disruption of a main-sequence star does not adequately explain UHECR observations.
Ref.~\cite{Zhang:2017hom} considers white dwarfs (WDs) disrupted by an intermediate-mass black hole (IMBH). 
They found that while common carbon-oxygen WDs (CO-WDs) poorly fit the spectrum, they could still account for the UHECR composition. 
In contrast, oxygen-neon-magnesium WDs (ONeMg-WDs) better match the Auger spectrum and composition, with an initial mass composition of $f_{\rm O} = 0.12$, $f_{\rm Ne} = 0.76$, and $f_{\rm Mg} = 0.12$.  
However, ONeMg-WDs are much rarer than CO-WDs, with a roughly $\sim  1 / 30$ ratio compared to CO-WDs.
However, CR nuclei accelerated in the internal shock, prompt emission region, cannot survive, unlike those in low-luminosity TDEs~\cite{Zhang:2017hom}.

\subsection{Active galactic nuclei}
Active galactic nuclei (AGN) are considered promising sources of UHECRs.
Radio-loud AGN (or jetted AGN) have multiple sites of efficient electron and ion acceleration, including the black hole vicinity (several $R_g$), inner jet, kiloparsec-scale jet, and jet lobes (see Ref.~\cite{2009arXiv0909.2576M}). 

Blazars, a subclass of AGNs, are particularly powerful high-energy sources where CR nuclei can be accelerated to ultrahigh energies~\cite{Murase:2014foa, Rodrigues:2018bjg}. 
In the case of BL Lacs, UHECR nuclei are efficiently accelerated; however, in FSRQs, the accelerated nuclei tend to disintegrate due to the intense radiation field in the vicinity of the jet~\cite{Rodrigues:2018bjg}.
FSRQs are also thought to be excellent sources of UHE neutrinos. 
High-energy neutrino emissions from blazars are constrained by experiments such as Auger, particularly for optimistic models (e.g., $s = 2.3$). However, this is highly model-dependent due to uncertainties in spectral indices~\cite{Murase:2014foa}.

In recent years, radio galaxies have received great attention as UHECR sources.
Acceleration processes in kiloparsec-scale jets and lobes have been studied in greater detail, particularly through advanced numerical simulations. For relativistic jets, one intriguing process is one-shot acceleration, where CRs experience a dramatic energy boost of $\sim \Gamma^2$ if the angle between the initial and final directions of flight exceeds $\pi/2$, accelerating particles to extremely high energies in a single interaction~\cite{Mbarek:2019glq}. 
The shear acceleration mechanisms have been proposed as a viable alternative for accelerating UHECRs in the jet-cocoon system. The discrete shear acceleration process, a type of Fermi acceleration mechanism, results in CRs being accelerated and exhibiting a power-law energy spectrum. The escaping spectrum of CRs is hard, $dL_E/dE \propto E^\alpha$, with $\alpha=0-1$. The maximum attainable energy is estimated as $E_{i, \rm~max} \approx \zeta e Z_i B_{\rm coc}l_{\rm coc}^{1/2} R_{\rm jet}^{1/2} \Gamma_{\rm jet} \beta_{\rm jet} \sim 1.6 Z_i\rm~EeV$, where $\zeta \sim 2.2$.
Hydrodynamic flow through hydromagnetic flux tubes in the backflows of radio galaxy lobes provides a conducive environment for accelerating CRs to ultrahigh energies~\cite{Bell:2019nnf}. This acceleration process can be understood through two overlapping mechanisms: first-order Fermi acceleration driven by shocks and second-order Fermi acceleration facilitated by strong turbulence. Within a single flux tube, the maximum energy attainable by cosmic rays at a shock is approximately equivalent to the Hillas energy limit $E_{\rm max} \approx ZuBl$, where $u$ is the flow velocity along the flux tube and $l$ is the width of the flux tube.

Concerning the composition of UHECRs, their origin and the processes involved in their acceleration remain areas of active research. 
The Espresso and Shear acceleration mechanisms are known to play a role in the acceleration of low-energy Galactic CRs.
Ref.~\cite{Kimura:2017ubz} considers injection at TeV energies and further assumes that the fraction of intermediate- and heavy-mass nuclei is enhanced by a factor of $\sim 3$. However, if the injection occurs at the same rigidity, it becomes challenging to simultaneously fit both the spectrum and the composition.

It is believed that different acceleration sites within AGNs could contribute to the varying compositions of UHECRs detected on Earth. 
The nature of particle disintegration and the specific interactions with the radiation field in the jet regions could play key roles in shaping the observed UHECR spectrum. 

\begin{figure}
    \centering
\includegraphics[width=\linewidth]{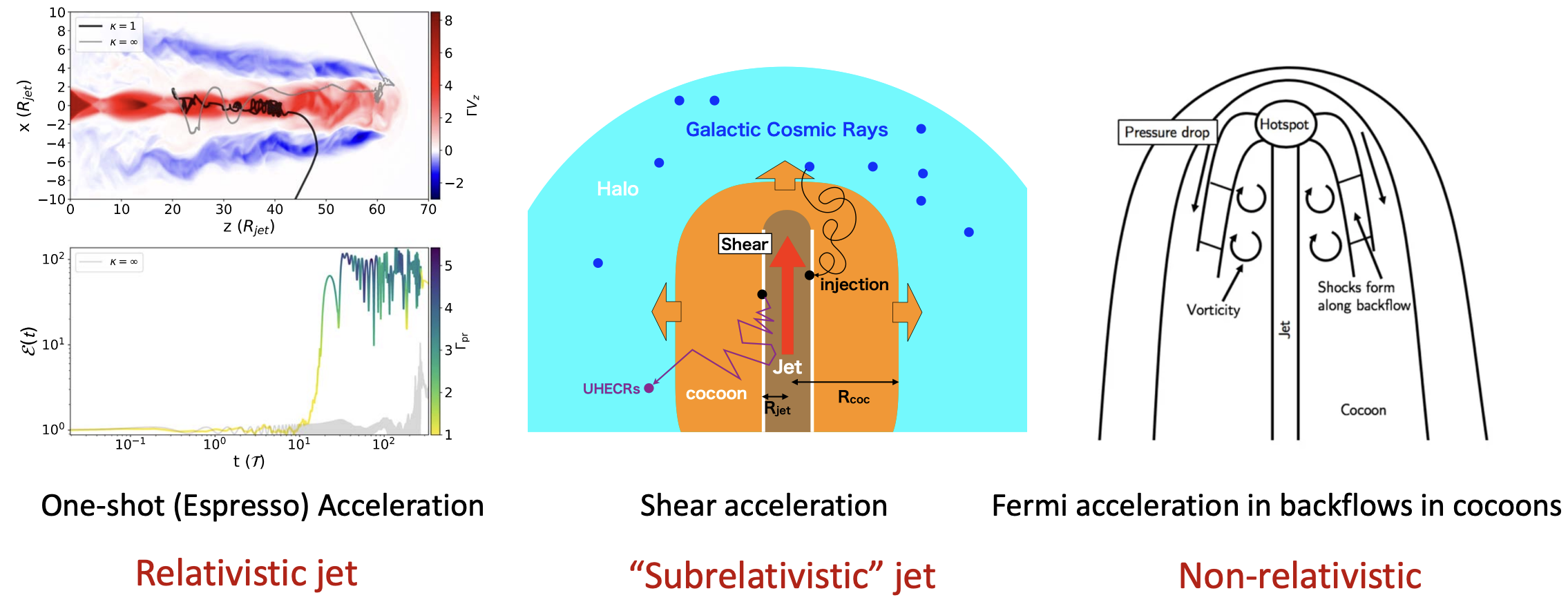}
    \caption{Various acceleration processes in AGN jets. Figure adapted from Ref.~\cite{Mbarek:2022nat, Kimura:2017ubz, Bell:2019nnf}.}
    \label{fig:AGN-acceleration}
\end{figure}

\section{Secondary gamma-rays and neutrinos from UHECR sources}
\begin{figure}
    \centering
    \subfigure[]{
    \includegraphics[width=0.48\linewidth]{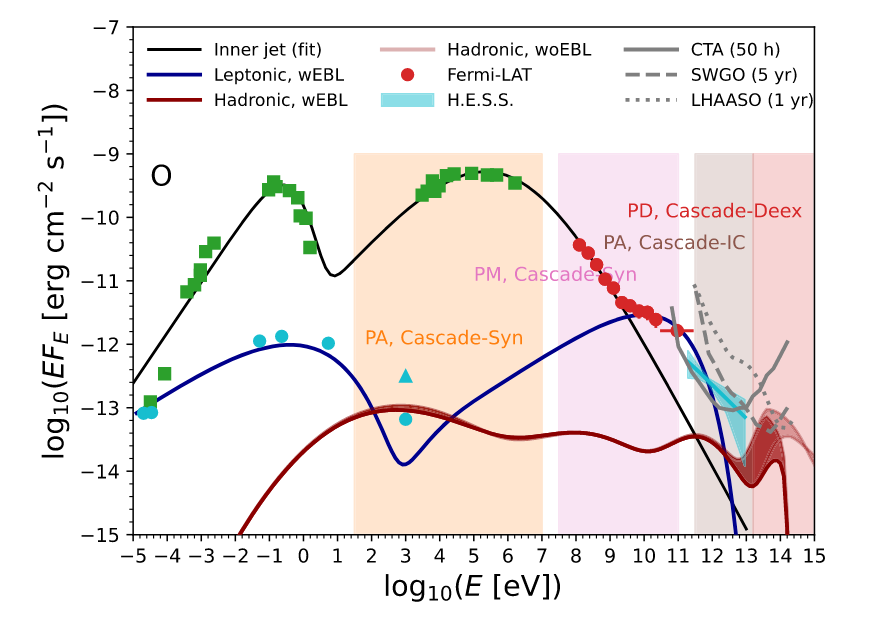}\label{fig:CenA-a}
    }
    \subfigure[]{
    \includegraphics[width=0.45\linewidth]{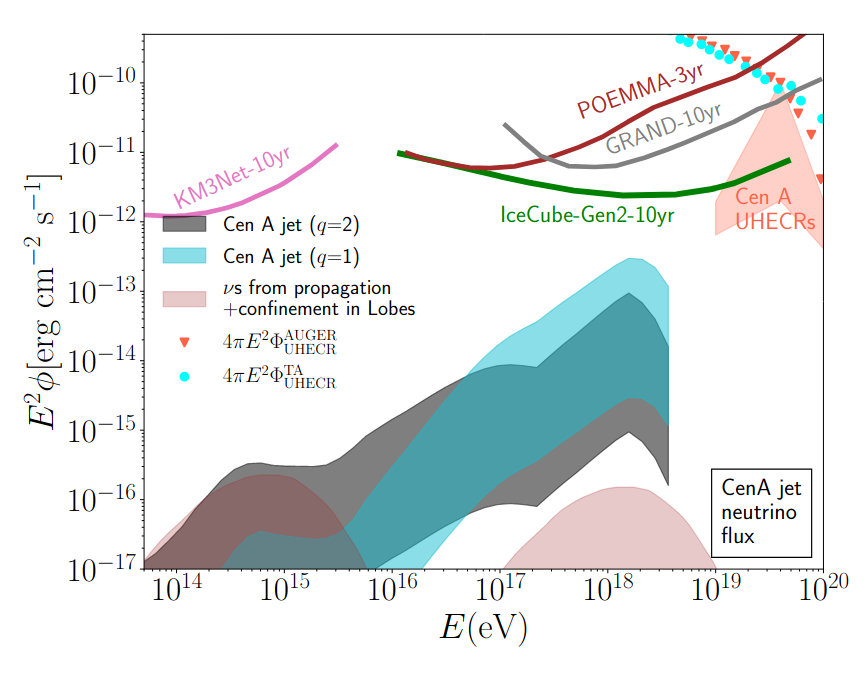}\label{fig:CenA-b}
    }
    \caption{Left panel: Predicted multiwavelength SED from the kiloparsec-scale jet after the injection of primary high-energy electrons (solid blue lines) and UHECR nuclei (solid red lines). Figure adapted from Ref.~\cite{Zhang:2023ewt}. Right panel: Expected neutrino flux from UHECRs accelerated in Cen A. Figure adapted from Ref.~\cite{Mbarek:2022nat}.}
    \label{fig:CenA}
\end{figure}

Secondary gamma-rays and neutrinos produced inside UHECR sources are powerful tools for exploring UHECR sources. 
The sources where UHECR nuclei can survive would be optically thin to high-energy gamma rays.
Here, we discuss observations of the radio galaxy Centaurus A (Cen A). UHECRs could be accelerated in the large-scale jet of Cen A, where interactions with ambient matter and photons lead to the production of secondary neutrinos and gamma rays.
In Fig.~\ref{fig:CenA-a}, the detection of de-excitation gamma rays is possible with the next-generation ground-based gamma-ray telescopes. 
High-energy neutrinos from the Cen A jet are shown in Fig.~\ref{fig:CenA-b}.

\section{Summary and perspective}
The study of UHECRs has progressed significantly in the Auger and TA era, offering new insights into their sources, composition, and anisotropy. 
Sources related to massive stellar deaths, including HL and LL GRBs, engine-driven supernovae, pulsars, and pulsar-driven transients, can provide enhanced intermediate- and heavy-mass nuclei to explain the observed composition. While high-energy neutrino observations constrain HL GRBs, LL GRBs and engine-driven supernovae offer more promising conditions for UHECR production.

Compact binary mergers (NS-NS and BH-NS) may also contribute, especially to the production of ultraheavy UHECRs via r-processes in neutron-rich environments. Pulsars, magnetars, and TDEs are additional potential sources of UHECRs, with accelerated particles originating from the surfaces of neutron stars or the jets of disrupted stars near supermassive black holes.

AGN, particularly radio galaxies, are promising candidates, with CR nuclei accelerated in large-scale jets and lobes. However, the interactions in AGN jets and radiation fields remain complex and are still under study.
Acceleration processes in kiloparsec-scale jets and lobes of radio galaxies are promising mechanisms for accelerating UHECRs.

Looking ahead, next-generation observatories like AugerPrime~\cite{Castellina:2019irv} will provide more detailed data, improving mass resolution and enabling the identification of new UHECR events. These future observations will help refine our understanding of UHECR origins and acceleration mechanisms, offering crucial multi-messenger insights into this cosmic mystery.

\section{Acknowledgments}
We extend our gratitude to our colleagues for their valuable contributions to advancing our understanding of UHECR sources. In particular, we thank Kohta Murase, Peter Mészáros, Shigeo Kimura, Shunsaku Horiuchi, and others for their insightful input and collaboration.

\bibliographystyle{unsrt}
\bibliography{bzhang}


\end{document}